\documentclass[12pt]{amsart}

\usepackage{graphicx}
\usepackage{amsmath}
\usepackage{amssymb}
\usepackage{amsthm}
\usepackage[margin=1in, top=1.5in]{geometry}
\usepackage{booktabs}

\usepackage[slantedGreek]{mathptmx}
\usepackage{url}
\usepackage{bbm}
\usepackage{accents}
\usepackage[nomarkers,figuresonly]{endfloat}

\newtheorem*{theorem*}{Theorem}

\newcommand{\Var}{\mbox{\rm Var}}

\newcommand{\alphat}{\tilde{\alpha}}
\newcommand{\betat}{\tilde{\beta}}

\title[ASPP in heterogeneous populations]{Psychological dimension of adaptive trading in cryptocurrency  markets}
\author{Misha Perepelitsa}

\thanks{Email: mperepel@central.uh.edu}

\date{\today}
\address{Misha Perepelitsa \\
mperepel@central.uh.edu\, 
University of Houston\\
PGH 631\\
4800 Calhoun Rd. \\
Houston, TX\\
USA}

\date{}

\begin{document}

\begin{abstract}
In this paper we extend the analysis of an agent-based model for adaptive trading, called asynchronous stochastic price pump (ASPP) introduced by Perepelitsa and Timofeyev (2019), to the model with heterogeneous distribution of psychological parameters of speculative optimism and pessimism  across the population of traders. We show that the new model has a range of qualitatively different dynamics when the correlation between those factors ranges from low negative to large positive values. A statistical parameter estimation suggests a heterogeneous ASPP with negative correlation as a model of price variations of Bitcoin.

\end{abstract}


\maketitle

\section{Introduction}
In \cite{PerTim19, PerTim21} we introduced a model for adaptive speculative trading, called Asynchronous Stochastic Price Pump (ASPP), and argued that the model might account for price dynamics of commodities that do not allow for computation of a fundamental value attached to it, such as cryptocurrencies. 

In this model, each agent re-balances its  investment portfolio, moving cash from savings to stocks or vice-versa depending on how well the price dynamics matches the agent's expectation. The difference in expectations creates the demand and supply of stocks and the new price is set to balance that relation.

Two important features  distinguish ASPP from other agent-based models for speculative trading. The first is that the model has an equilibrium state  in which price e doesn't change and neither the investment portfolios of agents. Since the model in completely endogenous, this state would be the only ``rational'' outcome. However, if the portfolios are out of balance, even by a small degree, the trading results in a non-trivial, divergent price dynamics, see Figure \ref{fig:inst_growth}. That  is, the rational, equilibrium state is unstable, and the instabilities grow into a price bubble.

The second property is that the behavior of agents is adaptive. Adaptation is ubiquitous in many aspects of human life including the speculative decision making (decision making under risk). In ASPP model, an agent will decrease the investment in stocks whenever the market under-performs as measured by the agent's stock-to-bond ratio, and will increase the investment otherwise. The adaptation is implemented in ASPP through the levels of emotional response of the agent to the performance of its portfolio: the levels of greed and fear. They are expressed as positive scalar parameters, expressing the intensity of the emotional response. In the model, agents use them as multipliers for the anticipated investment ratios.
In this context, the price bubble created by such process can been seen as the result of adaptation of agents to the ever changing ``environment'' that they create.

The results in \cite{PerTim19, PerTim21} were obtained  under a simplifying assumption that all agents have the same levels of greed (investor optimism) and fear (investor pessimism). Thus, the model was completely specified by only two psychological parameters $(\alpha,\beta),$ that enter into formulas for the return, the volatility and the risk of a crash. In this paper we remove that artificial condition from the ASPP model and augment the analysis by considering a heterogeneous population of traders with varying levels of greed and  fear. Our main goal is in discovering how the dynamics of ASPP depends on the correlation between $\alpha$ and $\beta$ across the population. We use direct, agent-based simulation to show the model has a range of  qualitatively different dynamics when the correlation ranges from low negative to large positive values. The differences are the returns, the risk of a crash,the character of wealth re-distribution, and the market performance of agents with different levels of psychological response.

From the point of view of applications, one might be interested in the question  which heterogeneous ASPP model is a better model for cryptocurrencies such as Bitcoin. We address this question by an indirect statistical test, by comparing the short term correlation in the temporal variations $\alpha$ and $\beta,$ obtained on the basis of the homogeneous ASPP for Bitcoin, Apple stock and index fund S\&P500 price series with the dynamics generated by heterogeneous ASPP with correlation of greed and fear in population ranging from -1 to 1. The comparison suggest that ASPP with negatively correlated values of the psychological parameters is  qualitatively better fit as a model for Bitcoin dynamics.

\subsection{Literature Review}
Bubbles in financial markets is well researched area.  Large price fluctuations in stock prices can be explained by the presence of noisy traders that imitate other traders and look at the market trends (Bak et al. 1997,  Levy et al. 1994, 1995, 2000, Lux 1995, 1998, Lux and Marchesi 1999, 2000). Price fluctuations can arise in the process of ``genetic'' evolution of trading strategies, when traders constantly adopt and modify existing ones, selecting better performing strategies, as in the model by  Arthur et al. (1997). Price bubbles may arise from rational expectations of traders and be consistent with efficient market hypothesis. This type of price dynamics was introduced by Mandelbrot (1966) and extended by Blanchard (1979),  Blanchard and Watson (1982). Within this framework, Johansen et al. (1999a, 1999b, 2000), Sornette (2003) constructed realistic non-linear models of price series that are characterized by price bubbles and crashes. 

Agent based modeling is by now a well-developed and widely used methodology. The review of agent based modeling in economics and finance can be found in Chakraborti et al. (2011), Chen et al. (2012), Iori and Porter (2014). There are few papers that apply the agent-based modeling to the cryptocurrncies trading. 
On this topic we would like to cite works Cocco et al. (2019, 2019b) who analyze the price dynamics generated by the traders using the genetic algorithms, technical analysis, or  random strategies. 

The significance of adaptive economic behavior  was stressed  by many authors in different contexts. In strategic decision making  it was advanced  by Cross (1983), in models of learning in games by Fudenberg and Levine (1998), and reinforcement learning models by Roth and Erev (1995), to mention just a few contributions.  Other psychological factors greatly influence trading activities in the markets  such as herding effects or follow-the-leader strategies. These were discussed, for example, by Lux (2005), Tedeschi (2012)  and Garcia and Schweitzer (2015)

There are several studies that use the econometric statistical tool to the analysis of Bitcoin time series. They can be found in Liu and Tsyvinksi (2018) and Makarov and Schoar (2020).

\section{Model parameters estimation from data sets}
For values of $\alpha,\beta$  close to 1, in \cite{PerTim19, PerTim21} we 
we showed that the logarithm of the the return, $R(t),$ of a price series of an ASPP, in a long run, is  distributed according to a normal distribution
\begin{equation}
\label{eq:return2}
\ln(R(t))\in  \mathcal{N}(\ln r,\,\sigma)
\end{equation}
with the formulas the rate of return $r$ (geometric mean) and the volatility $\sigma$ expressed by formulas
\begin{equation}
\label{eq:r_sigma}
r{}={}\left(\frac{\alpha}{\beta}\right)^{\frac{1}{\gamma}},\quad \sigma{}={}c_0(\alpha\beta-1),
\end{equation}
where $\gamma{}={}\frac{2N}{m}.$ In this formula, $N$ is the number of traders, $m$ in the number ``active traders,'' that is, the traders participating in buying and selling per day. $c_0$ is a positive constant. For a reference, we note that in \cite{PerTim19, PerTim21}, the inverse of $\betat$ was used instead in  formulas for $r$ and $\sigma.$

Let $P(t)$ be a time series of a unit of  traded commodity (closing price at the end of day $t$). In the examples below we consider daily prices of a Bitcoin, Apple stock and S\&P500 index for the last 8 years, ending in March 2021\footnote{Price quotes were obtained from https://finance.yahoo.com}. 
We'll assume that the short term fluctuations in the prices are well approximated by an  ASPP model with 
some values of $\alpha$ and $\beta$  which we'd like to estimate the time series. Using a moving statistics over intervals of $\tau=7,$ $14,$ or $21$ days, we compute  the moving averages and the standard deviations: 
\[
\ln r(t) = \tau^{-1}\sum_{k=1}^\tau \ln\left(P(t+k)/P(t+k-1)\right),
\]
and
\[
\sigma^2(t) {}={}(\tau-1)^{-1}\sum_{k=1}^\tau \left[\ln\left(P(t+k)/P(t+k-1)\right){}-{}\ln r(t)\right]^2,\]
for each day $t=1..M,$ when the information is available (in our examples, in the last  8 years). 

Given $r(t)$ and $\sigma(t),$ equations \eqref{eq:r_sigma} can be inverted to obtain values  $\alpha(t)$ and $\beta(t).$ To avoid the confusion and to distinguish between the values of $\alpha$ and $\beta$ defining an ASPP model and their statistical estimates over a short term window $\tau$ from a time series of the price of a commodity, we introduce notation $\alphat$ and $\betat$ for the latter. 
In particular, we will not assign the same psychological interpretation of greed and fear to these new parameters, as we did with $\alpha$ and $\beta.$
Thus, we obtain
\begin{equation}
\label{eq:ln_a_b}
\ln\alphat {}={}\frac{\gamma}{2}\ln r{}+{}\frac{1}{2}\ln\left(\frac{\sigma+c_0}{c_0}\right),
\quad
\ln\betat {}={} -\frac{\gamma}{2}\ln r{}+{}\frac{1}{2}\ln\left(\frac{\sigma+c_0}{c_0}\right).
\end{equation}
The formulas involve model parameters $\gamma$ and $c_0,$ the values of which are difficult to estimate, and thus, direct estimation of values of $\alphat,$ $\betat$ seems to be problematic. To address this difficulty we perform  data analysis for  a range of $\gamma$'s, varying from a low to high.  The indeterminacy in $c_0$ will be resolved by selecting its value so that the changes in $\ln(\sigma+c_0)/c_0$ are of order of changes of $\gamma\ln r.$ We use this approach to compare the correlation  between $\alphat$ and $\betat$  for price time series of different commodities, for the given range of $\gamma$'s.

\begin{figure}
\centering
\includegraphics[width=0.7\columnwidth]{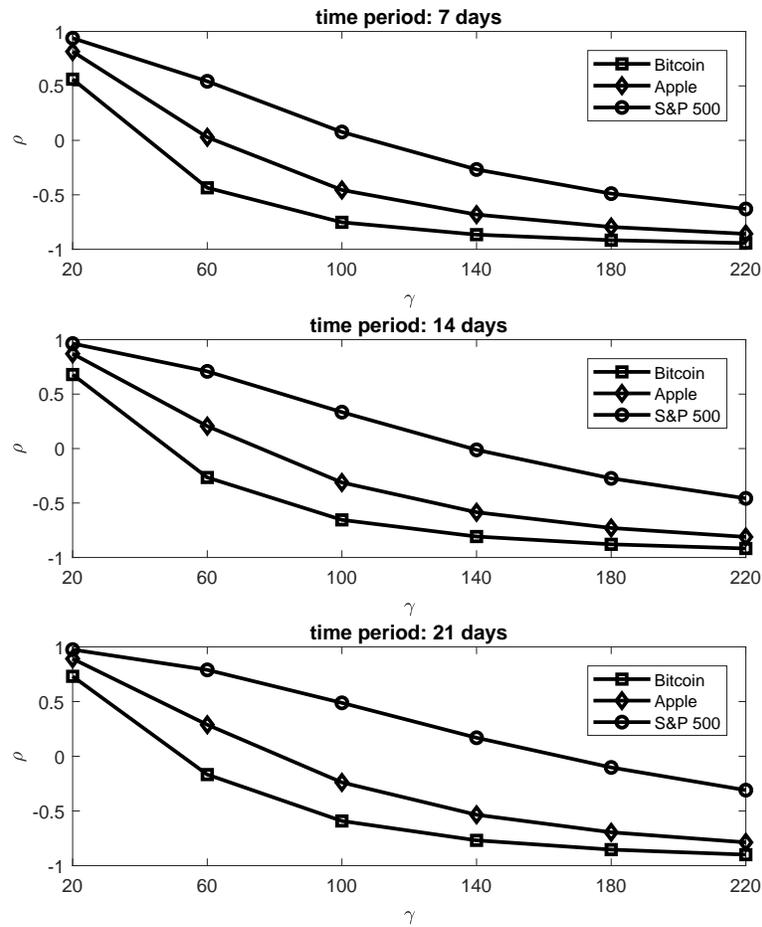}
\caption{Correlation coefficient $\rho$ between $\ln\alphat(t)$ and $\ln\betat(t)$ for Bitcoin, Apple stock and index fund S\&P500. The range of parameter $\gamma\in[20,220],$ see \eqref{eq:ln_a_b}. The mean return and volatility are computed over periods of  $\tau{}=7, 14$  and $21$ days.}
\label{fig:corr1}
\end{figure}

The results are presented in Figure \ref{fig:corr1}. It shows the correlations
between $\ln\alphat(t)$ and $\ln\betat(t),$ obtained in this way  for different values of $\gamma.$ Notice from \eqref{eq:ln_a_b}, that selecting a low value of $\gamma$ results in a correlation close to 1, and selecting a large value, brings the correlation closer to -1. This is reflected in the plots at values of $\gamma=20,$ and $220,$ respectively. The non-trivial information about correlations is contained in the intermediate values of $\gamma.$ 

The key feature of the plots is a consistent order in the correlations for Bitcoin (lowest), Apple (intermediate) and S\&P500 (largest), for all values of $\gamma.$  Moreover, the correlation is negative for Bitcoin in most cases, and positive for the index fund S\&P500 in half of the cases. 

\begin{figure}
\centering
\includegraphics[width=\columnwidth]{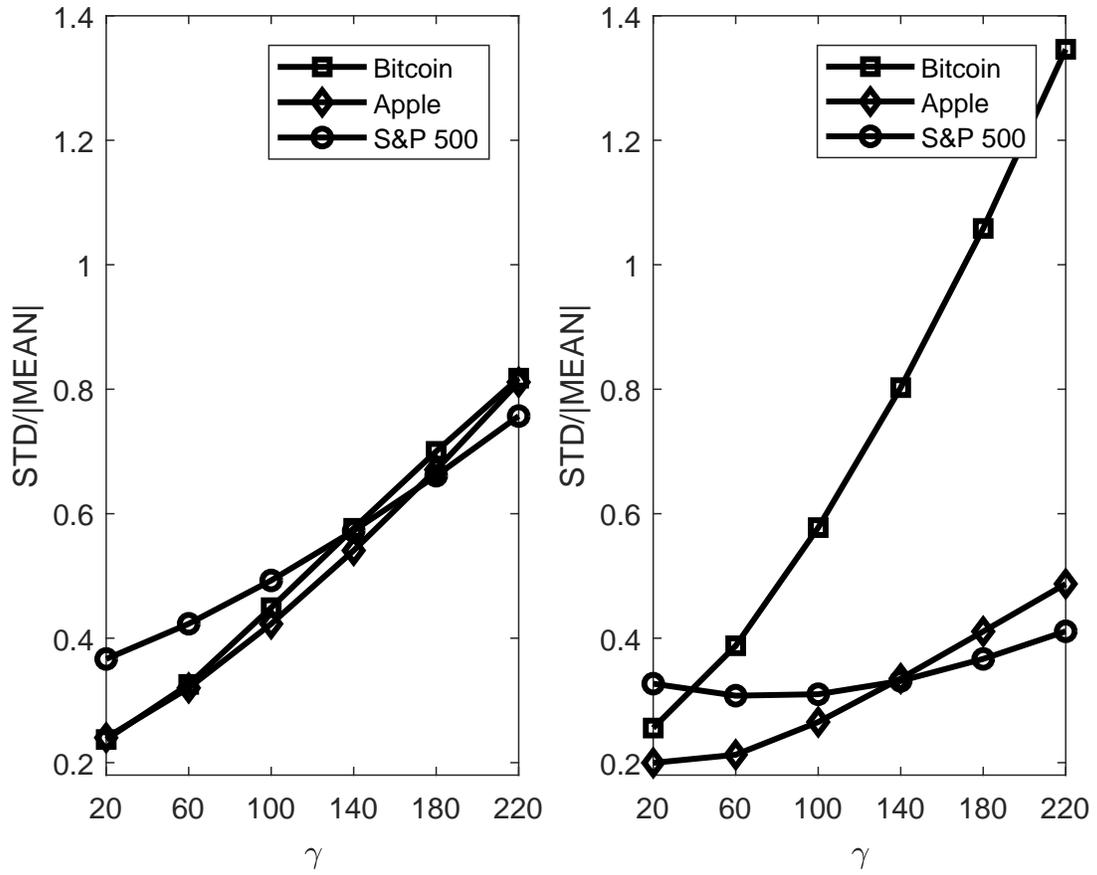}
\caption{Coefficient of variance (standard deviation/$|\textrm{mean}|$) for $\ln\alphat(t)$ (left) and $\ln\betat(t)$ (right) as a function of parameter $\gamma,$ see \eqref{eq:ln_a_b}.}
\label{fig:coef_variation}
\end{figure}

Figure \ref{fig:coef_variation} shows the values of the coefficient of variation (standard deviation/mean) for $\ln\alphat$ (left) and $\ln\betat$ (right), for different values of $\gamma,$ computed used short term statistics over $\tau=21$ days.  Notice that all plots exhibit  appreciable changes in both parameters  during the time span of 10 years. The nature of these variations is an interesting question itself. It may result from all agents re-evaluating their emotional response  to the changes in wealth  when new information becomes available, thus, incorporating exogenous effects. This is likely to apply to Apple stock and S\&P500 dynamics.  Alternatively, the variations can be generated endogenously, for example due to variations in $\alpha$'s and $\beta$'s across the population of agents, and it might be relevant in Bitcoin case, as it lacks any fundamental value anchors. Following the hypothesis that was expressed in \cite{PerTim21}, that ASPP model is close proxy for Bitcoin price variations, in the present work we focus on studying  ASPP models driven by heterogeneous population of traders. That is, we will assume that the greed and fear factors $\alpha$ and $\beta$ vary across the population, and compare different ASPP models by changing the correlation (across the population) between $\ln\alpha$ and $\ln\beta$ from negative to positive. The detail of the new model, called heterogeneous ASPP, are presented in the following sections. 

Here, we would like to compare the statistics for the correlation of $\ln\alphat(t)$ and $\ln\beta(t)$ in \eqref{eq:ln_a_b} on a simulated time series of heterogeneous ASPP  with the correlation parameter  between greed and fear ranging over values $r=-1,$ $0.5,$ and $1.$

\begin{figure}
\centering
\includegraphics[width=0.7\columnwidth]{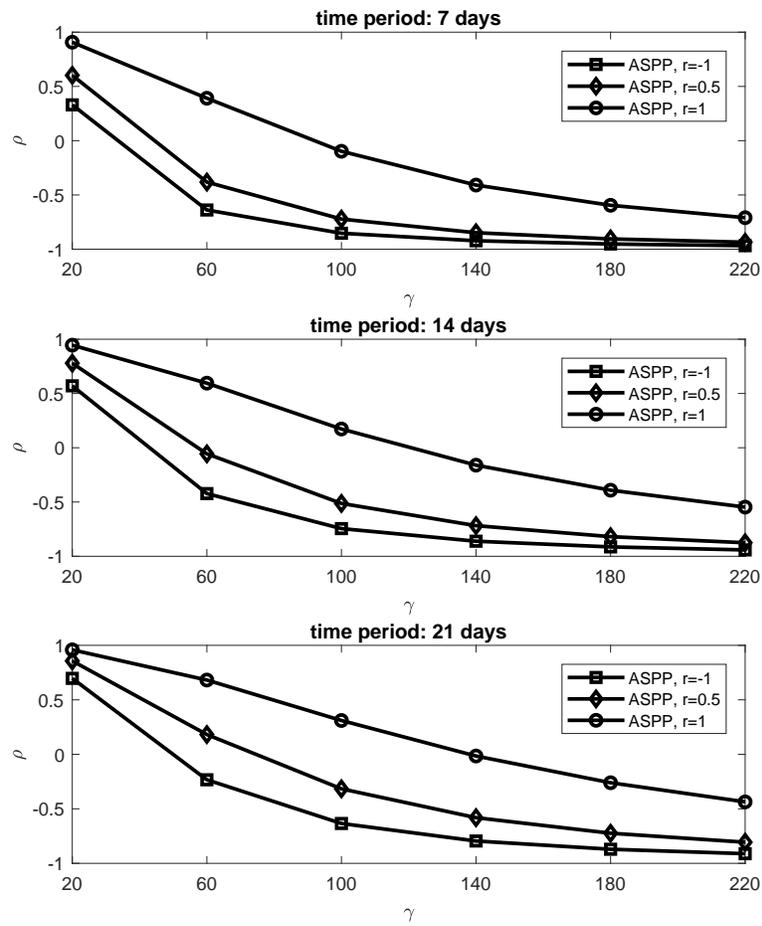}
\caption{Correlation coefficient $\rho$ between $\ln\alphat(t)$ and $\ln\betat(t)$ for price series generated by heterogeneous ASPP model  with values of the correlation coefficient $r=-1,0.5, 1.$ The range of parameter $\gamma\in[20,220],$ see \eqref{eq:ln_a_b}. The mean return and volatility are computed over periods of  $\tau{}=7, 14$  and $21$ days.}
\label{fig:corr2}
\end{figure}

\begin{figure}
\centering
\includegraphics[width=\columnwidth]{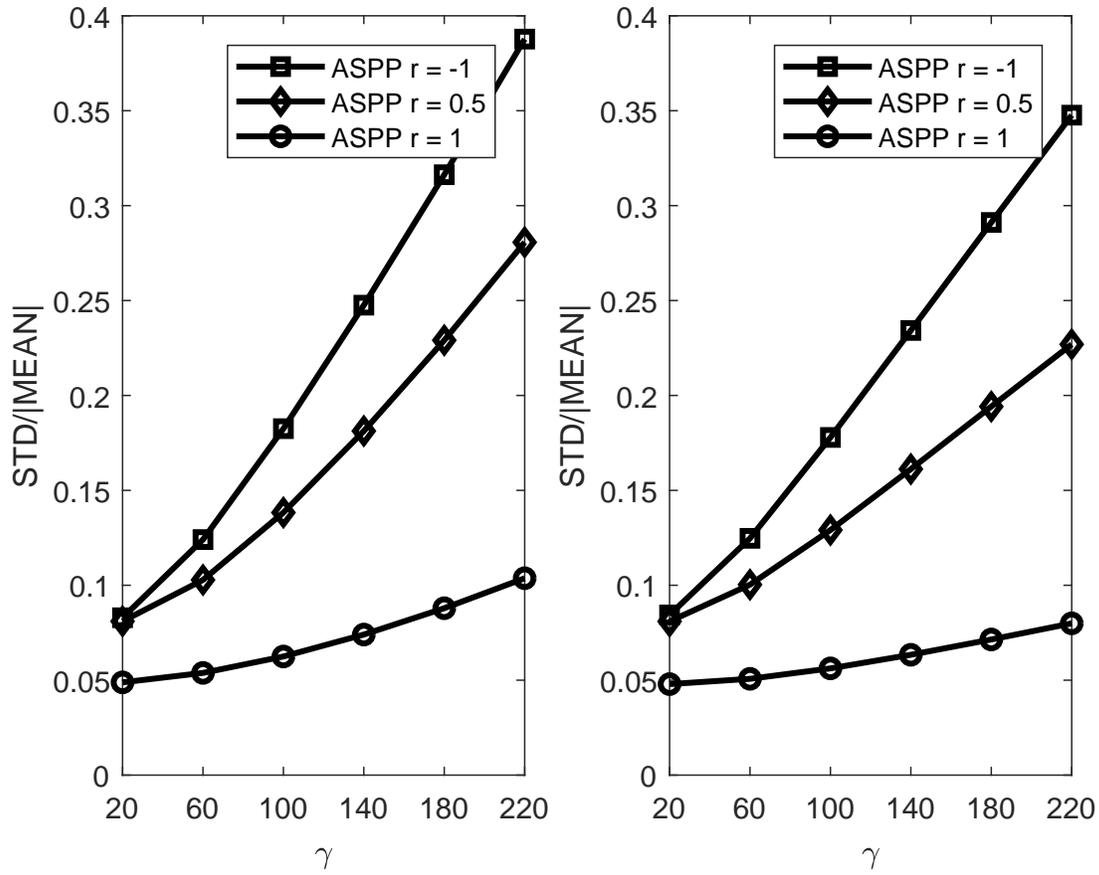}
\caption{Coefficient of variance for $\ln\alphat(t)$ (left) and $\ln\betat(t)$ (right) for a price series generated by heterogeneous ASPP model  with values of the correlation coefficient $r=-1,0.5, 1.$}
\label{fig:coef_variation2}
\end{figure}

Figure \ref{fig:corr2} show the correlation plots which qualitatively reproduces the features of data sets of Bitcoin, Apple and S\&P500 from Figure \ref{fig:corr1}. Figure \ref{fig:coef_variation2} shows that the coefficient of variation for heterogeneous ASPPs is of the order of magnitude of the coefficient of variation from Figure \ref{fig:coef_variation}.

The above analysis that indicates that short term price variation of Bitcoin are best described by a heterogeneous ASPP with negative correlation between greed and fear factors. Whereas for the index fund S\&P500 the correlation is positive or zero. Thus, postulating that ASPP is a valid model for Bitcoin for longer times, short term statistics  can be used to make to predictions for long term price fluctuations through simulation of an ASPP model.

\section{ASPP in heterogeneous populations: the model}
\label{sec:model}
For a mulit-agent model, we populate the market with many copies of agents considered above  and let them trade anonymously, by submitting their buy-sell orders to a market maker.  Thus,
 we consider a set of $N$ agents, described by their portfolios $(s_i,b_i),$ $i=1\ldots N,$ of dollar values of a stock and cash accounts, and let $k_i$ stand for agent $i$ target stock-to-bond ratio. $P_0$ will denote a current price per share and $P$ the new price determined by agents' demand. Let $\{i_l\,:\,l=1\ldots m\}$  be the set of  ``active'' agents, i.e. the ones setting the new price. The set of active traders is determined each trading period by a random draw from the population.

If $x_{i_l}$ is the dollar amount that agent $i_l$ wants to invest in stock, then
\[
\frac{\frac{P}{P_0}s_{i_l}+x_{i_l}}{b_{i_l}-x_{i_l}}{}={}k_{i_l}.
\] 
 The demand-supply balance is a simple market clearance condition
\[
\sum_{l=1}^m x_{i_l}{}={}0,
\]
which can be solved for $P:$
\begin{equation}
\label{eq:price}
\frac{P}{P_0}{}={}\left( \sum_{l=1}^m\frac{k_{i_l}b_{i_l}}{1+k_{i_l}}\right)\left(\sum_{l=1}^m\frac{s_{i_l}}{1+k_{i_l}}\right)^{-1}.
\end{equation}
Once the price is set, agents move corresponding amounts between cash and stock accounts, re-balancing their portfolios. 

Following that, the active agents update their expected stock-to-bond ratios from current $k_i$ to a new value $\hat{k}_i,$ using the agents specific greed ($\alpha_i\geq1$) or fear ($\beta_i\geq 1$) factors,
depending on whether an agent was selling or buying stocks:
\begin{equation}
\label{eq:feedback_prop}
\hat{k}_i{}={}\left\{
\begin{array}{ll}
\alpha_i k_i & \dfrac{Ps_i}{P_0b_i}>k_i,\\
\\
k_i & \dfrac{Ps_i}{P_0b_i}=k_i,\\
\\
\dfrac{k_i}{\beta_i} &\dfrac{Ps_i}{P_0b_i}<k_i.
\end{array}
\right.
\end{equation}
The trading session is repeated the following trading periods with new, randomly selected sets of active agents. The distribution of $\alpha$'s and $\beta$'s in a population defines a heterogeneous  model. In the examples, considered below, we assume that $(\ln\alpha,\ln\beta)$ is jointly Normal, cut to the range $\{\ln\alpha\geq0,\,\ln\beta\geq0\}.$ and denote by $r$ the correlation coefficient. This model extends the models considered in \cite{PerTim19, PerTim21} where it was assumed that all agents  have same greed and fear factors.

As was mentioned in the introduction, the model has an equilibrium state, when all agents have balanced portfolios, $s_i/b_i=k_i,$ $i=1,\ldots,N.$ In this case stock price doesn't grow, $P=P_0.$ When the parameters  are out of the equilibrium, even by a small degree,  the system exhibits non-trivial, divergent dynamics, such as shown in Figure \ref{fig:inst_growth}.

\begin{figure}
\centering
\includegraphics[width=0.7\columnwidth]{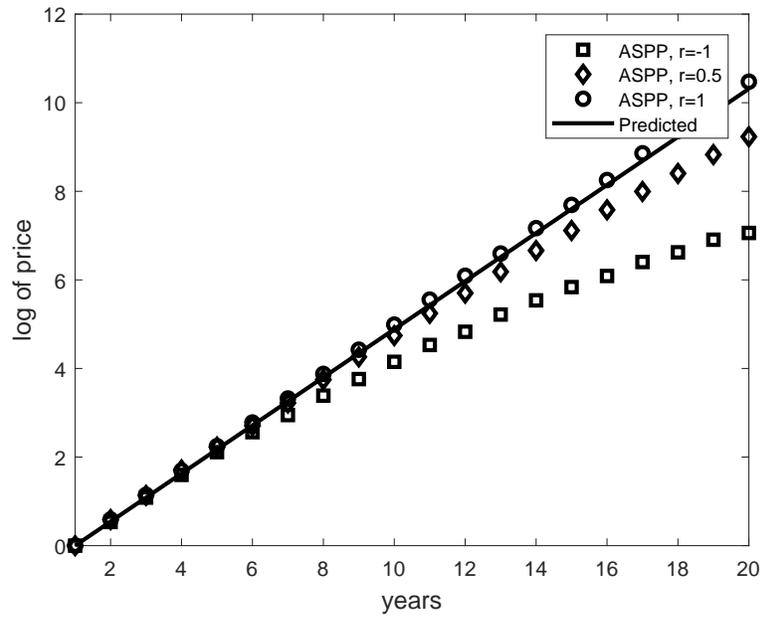}
\caption{Logarithm of the price generated by ASPP with the correlation parameter $r\in\{-1,0.5,1\}$ and the price predicted by formula \eqref{eq:r_sigma}. }
\label{fig:Price}
\end{figure}

As we will see in from the next section, the heterogeneous ASPP has time-varying mean returns, that are described by formulas \eqref{eq:r_sigma} of the homogeneous case only for certain transitional period, but then slowly decreasing, see Figure \ref{fig:Price}.
This can be explained by the character of wealth re-distribution during the process. 
As we will see in the next section, most of the cash in the system tends to be accumulated by agents with either lower greed and higher fear levels. A group of such agents has the mean value $\alpha/\beta$ less than than the average over all population. On the other hand price changes, given by \eqref{eq:price}, are stronger influenced by agent with more cash. Thus, during the lifetime of an ASPP price changes will driven by a lesser $\alpha/\beta$ ratio of group of agents with more cash, and with time the decrease accelerates.  The simulation show that the return rate of the homogeneous case in \eqref{eq:r_sigma} gives an upper bound for the rates of heterogeneous models.   

Simulations also show that the direction of correlation between greed and fear
 parameters is important for the performance of an ASPP.  For example, if all variation in $(\alpha,\beta)$ are along a line $\alpha/\beta{}={}const,$ then the price dynamics of such heterogeneous ASPP is identical with that of the single values of these parameters in the population, that is, the case considered in \cite{PerTim19, PerTim21}, see Figure \ref{fig:Price}.  On the other hand, the most deviation in the price occurs when the parameters are spread along lines $\alpha\beta{}={}const.$ 
More details on the price dynamics can be found in the next section.

\section{Positively vs. negatively correlated ASPPs}
\label{sec:correlations}

\begin{figure}
\centering
\includegraphics[width=0.7\columnwidth]{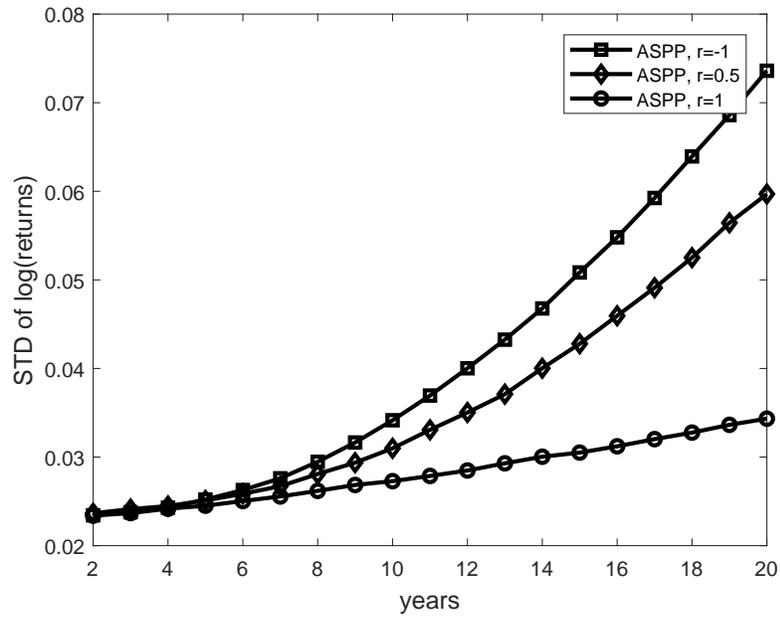}
\caption{Volatility of the returns generated by ASPP with the correlation parameter $r\in\{-1,0.5,1\}$.}
\label{fig:STD}
\end{figure}

\begin{figure}
\centering
\includegraphics[width=0.7\columnwidth]{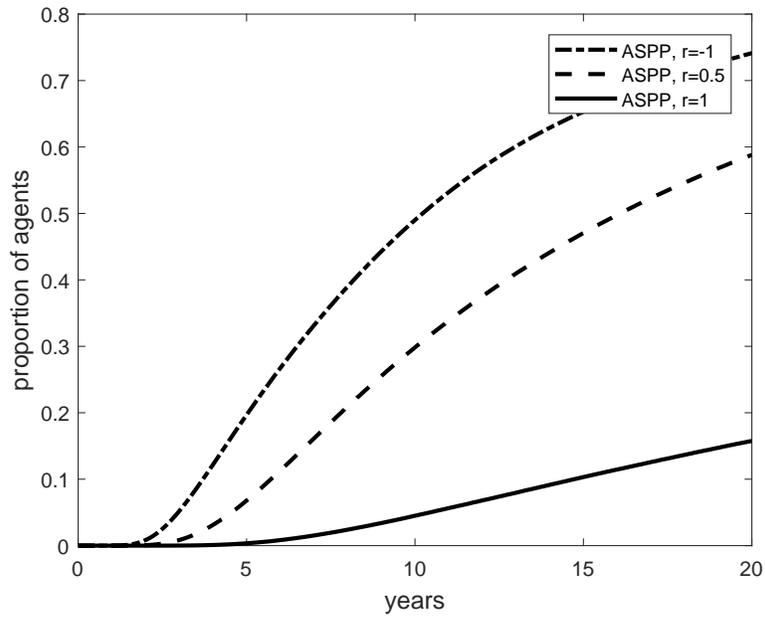}
\caption{Systematic risk. The risk of a crash is related to the proportion of agents with low (0.5) ratio of current cash to initial cash for ASPP with the correlation parameter $r\in\{-1,0.5,1\}$}
\label{fig:Risk}
\end{figure}

\begin{figure}
\centering
\includegraphics[width=0.7\columnwidth]{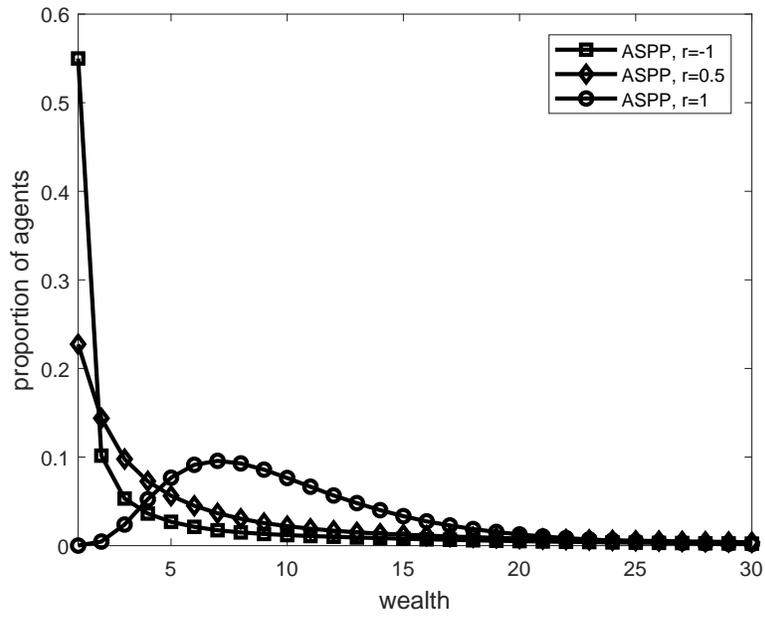}
\caption{Wealth (cash or bond) distribution in the population of agents at the end of the trading period of 20 years.}
\label{fig:Wealth}
\end{figure}

\begin{figure}
\centering
\includegraphics[width=0.7\columnwidth]{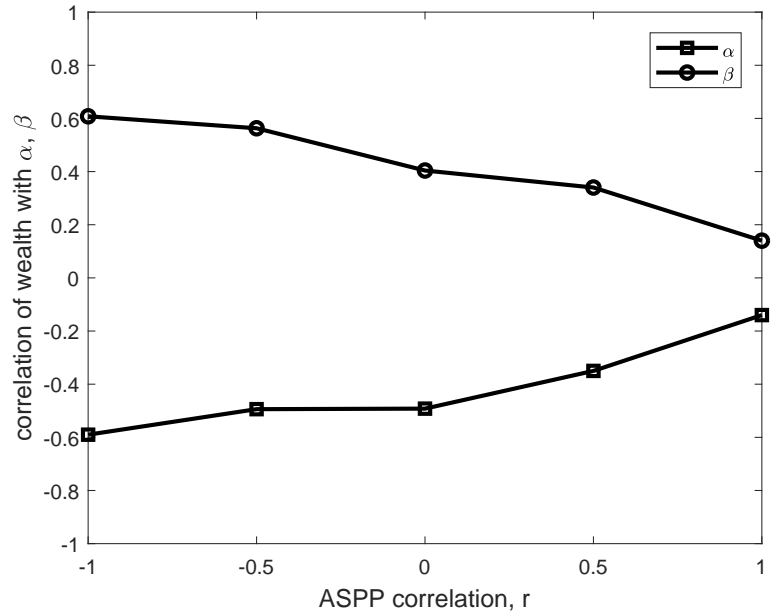}
\caption{Correlation between wealth (cash or bond) accumulated by agents and their levels of greed $\alpha,$ and fear $\beta$ for the correlation parameter of ASPP $r\in\{-1,-0.5, 0.5, 1\}.$  }
\label{fig:Wealth_alf_bet}
\end{figure}

We illustrate this behavior on a particular model setup, in which there is total of $N=500$ investors, with $m=20$ randomly chosen agents to participate in a daily trade. The values of  greed and fear factors for agents are randomly sampled from a jointly Normal distribution with the mean of $\ln\alpha=1.2,$ the mean of $\ln \beta=1.04,$  with equal variances: $\Var(\ln\alpha) = \Var(\ln\beta){}={}1.7\cdot10^{-4},$ and varying levels of the correlation coefficient $r\in[-1,1].$ The values of $(\ln\alpha,\ln\beta)$ are restricted to the model relevant range of $\{\ln\alpha\geq0,\ln\beta\geq0\},$ and the specific choice of variance guaranties that all sample points within 3 standard deviations from the mean are located in this range.  All agents start with \$10 in bonds initially, with target ratios $k_i=1,\,i=1..N,$ and starting price $P_0=1.$ Agent $i$ has
$
 \$(10k_i\,+\,\textrm{ random noise of order 0.1})
$
in stocks, where the small noise is added to perturb the unstable equilibrium state. We performed 100,000 path simulations of ASPP for daily trades for the duration of 20 years, for each level of correlation $r.$ The statistics of the simulations is presented in Figures \ref{fig:Price} through \ref{fig:Wealth_alf_bet}.

\subsection{Price and volatility}
Figure \ref{fig:Price} shows the logarithm of the stock price for the levels of correlation of greed and fear, $r=-1,0.5,1.$
The perfectly correlated case of $r=1$ shows a dynamics consistent with the predictions given by formulas \eqref{eq:r_sigma}, represented by a solid line in Figure \ref{fig:Price}. The volatility plots are sketched in Figure \ref{fig:STD}. For each day for simulated trades, it shows the sample standard deviation of the logarithm of the gross return, over 100,000 samples. The volatility exhibits a significant increase when the correlation coefficient $r$ decreases. This indicates that the process is non-stationary.  
\subsection{Systematic risk}
As in \cite{PerTim21}, we related the risk of a crash of ASPP to be related to the proportion of agents with large ratio of their current wealth (cash or bond) to the initial wealth. Here we set that proportion to an arbitrary level, say  of 0.5, and plot the dynamics of this risk proxy in Figure \ref{fig:Risk}. The plots show significant and monotone change in the risk with decreasing values of correlation $r.$. 
\subsection{Wealth redistribution}
The ASPP is a conservative model so that the total cash and number of stocks remain fixed. The trading results in re-distribution of wealth (cash or bonds owned by an agent), that initial was equal for all agents. Figure \ref{fig:Wealth} shows the histogram of wealth by the end of 20-year period. The character of wealth distribution is greatly affected by $r.$ For large negative correlations $(r=-1)$ the distribution is most unequal, with majority of agent being extremely low on cash, and with increasing $r$ the distribution  is more equalizing, showing that majority retain a fixed portion of the initial cash (\$10), so that effect is time dependent as reflected in Figure \ref{fig:Risk}.
\subsection{Best performing strategies}
An interesting question if market performance of an agent portfolio depends on a strategy the agent uses, that is, the greed and fear factors. We measured the correlation between the accumulated cash and $\alpha$ and $\beta$ parameters by the end of period of 20 years and plotted the corresponding values in Figure \ref{fig:Wealth_alf_bet}.
For negatively correlated ASPP there is strong negative correlation between the greed factor $\alpha$ and the wealth, and strong positive correlation between the fear factor $\beta$ and the wealth. Thus correlations decrease monotonically low values, of order 0.1 when the correlation increases to 1.  The correlation graphs appear to be somewhat symmetrical.
These properties explain why in the negatively correlated case the logarithm of the price of ASPP falls below the positively correlated. In the former case almost all of the cash in concentrated among agents with low $\alpha$ and high $\beta$ levels. This group has lower levels of $\alpha/\beta$ than the mean,  but has stronger influence of the price, thus generating lower returns.

\section{Conclusions}
We considered  a model of speculating trading which traders, adaptively, re-balance their portfolios and generate a price bubble. The novelty of our approach is in analyzing a population of traders characterized by a heterogeneous distribution of 
psychological factors of greed and fear that determine the price dynamics of the bubble. On the basis of the numerical simulation be obtained the dependence of such price characteristics as the mean return, the volatility, the risk of a crash and wealth distribution on the correlation parameter describing the population. The comparison of the model  with the time series of Bitcoin and other commodity suggested that levels of greed (investing optimism) and fear (investing pessimism) are negatively correlated across the population. 

A heterogeneous ASPP model with correlation $r=1$ appears to be an ``optimal'' price pump among  all other models with $r\in[-1,1].$ This model achieves the highest rate of return, the lowest volatility,  the lowest risk of a crash, and the least degree of wealth concentration.

Among the simplifying assumptions made in this analysis, was the endogenous model setup. The dynamics will depends in a non-trivial way when the in- and out-flow of capital allowed, as in the problem of investining ASPP. This will be considered in our future work. Some other extensions considered below.
\subsection{ASPP with proportional adjustment}
\begin{figure}
\includegraphics[width=0.7\columnwidth]{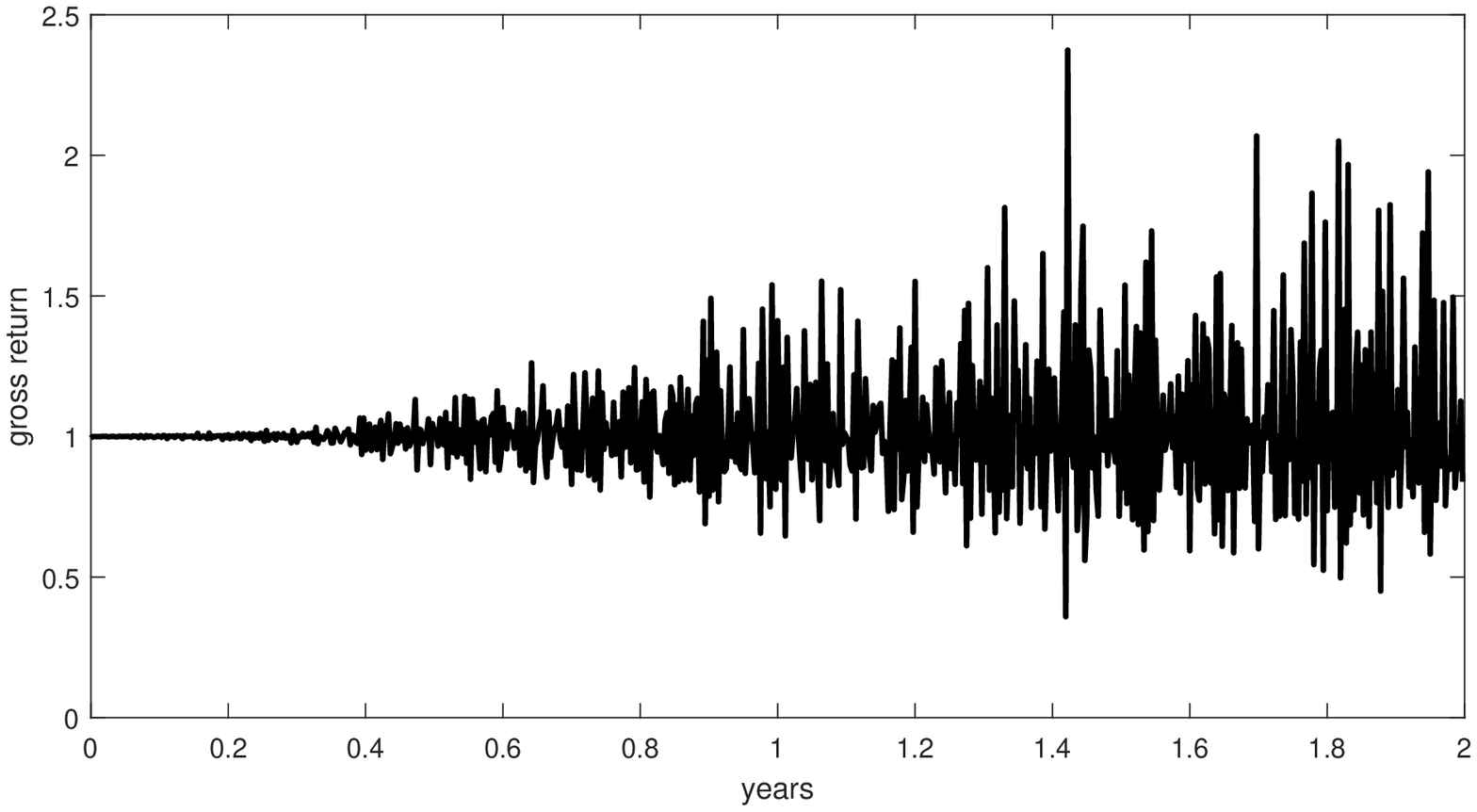}
\caption{Growth of instabilities in the returns in the price dynamics generated by an ASPP with proportional adjustment. Target stock-to-bond ratios are updated according to model \eqref{eq:feedback_prop1}.}
\label{fig:inst_growth}
\end{figure}

There are several variants of ASPP that add additional complexity for the model. We mention here some of  them.  The agents update the target portfolio ratios according to the rule
\begin{equation}
\label{eq:feedback_prop1}
\hat{k}_i{}={}\left\{
\begin{array}{ll}
(1 + \alpha_i |x_i|) k_i & \dfrac{Ps_i}{P_0b_i}>k_i,\\
\\
k_i & \dfrac{Ps_i}{P_0b_i}=k_i,\\
\\
\dfrac{k_i}{1+\beta_i |x_i|} &\dfrac{Ps_i}{P_0b_i}<k_i,
\end{array}
\right.
\end{equation}
$x_i$ is the dollar amount agent $i$ spend buying or selling. Figure \ref{fig:inst_growth} shows a simulated dynamics of return for this model of ASPP. Exponential  waiting times between trades for each agents might be too restrictive. One alternative is to have waiting times with memory and/or correlated with the price or changes in the price.  Systematic risk generated by ASPP discussed above can be included in the model to account for crashes. The analysis of these extensions will be performed in future works.

\end{document}